\documentclass[%
 aip,jap,
 amsmath,amssymb,
reprint,longbibliography,
]{revtex4-1}

\usepackage{graphicx}
\graphicspath{{Figures/}}
\usepackage[%
  colorlinks=true,
  urlcolor=blue,
  linkcolor=blue,
  citecolor=blue
]{hyperref}
\begin{document}

\title[Kateb \emph{et al.} disordered Ni$_{80}$Fe$_{20}$]{\color{blue}Effect of atomic ordering on the magnetic anisotropy of single crystal Ni$_\text{80}$Fe$_\text{20}$}

\author{Movaffaq Kateb}
 \email{mkk4@hi.is}
\affiliation{Science Institute, University of Iceland, Dunhaga 3, IS-107 Reykjavik, Iceland}
\author{Jon Tomas Gudmundsson}
 \email{tumi@hi.is}
\affiliation{Science Institute, University of Iceland, Dunhaga 3, IS-107 Reykjavik, Iceland}
\affiliation{Department of Space and Plasma Physics, School of Electrical Engineering and Computer Science, KTH--Royal Institute of Technology, SE-100 44, Stockholm, Sweden}
\author{Snorri Ingvarsson}%
 \email{sthi@hi.is}
\affiliation{Science Institute, University of Iceland, Dunhaga 3, IS-107 Reykjavik, Iceland}

\date{\today}

\begin{abstract}

We investigate the effect of atomic ordering on the magnetic anisotropy of Ni$_{80}$Fe$_{20}$~at.\% (Py). To this end, Py films were grown epitaxially on MgO (001) using dc magnetron sputtering (dcMS) and high power impulse magnetron sputtering (HiPIMS). Aside from twin boundaries observed in the latter case, both methods present high quality single crystals with cube-on-cube epitaxial relationship as verified by the polar mapping of important crystal planes. However, X-ray diffraction results indicate higher order for the dcMS deposited film towards $L1_2$ Ni$_3$Fe superlattice. This difference can be understood by the very high deposition rate of HiPIMS during each pulse which suppresses adatom mobility and ordering. We show that the dcMS deposited film presents biaxial anisotropy while HiPIMS deposition gives well defined uniaxial anisotropy. Thus, higher order achieved in the dcMS deposition behaves as predicted by magnetocrystalline anisotropy i.e.\ easy axis along the [111] direction that forced in the plane along the [110] direction due to shape anisotropy. The uniaxial behaviour in HiPIMS deposited film then can be explained by pair ordering or more recent localized composition non-uniformity theories. Further, we studied magnetoresistance of the films along the [100] directions using an extended van der Pauw method. We find that the electrical resistivities of the dcMS deposited film are lower than in their HiPIMS counterparts verifying the higher order in the dcMS case.
\end{abstract}

\pacs{77.55.Px, 75.30.Gw, 75.70.-i, 75.50.Bb, 81.15.Cd, 81.15.Kk, 87.64.Bx}

\keywords{HiPIMS, Epitaxy, NiFe, uniaxial anisotropy, order}

\maketitle 

\section{Introduction}

The phenomenon of \emph{uniaxial anisotropy} in permalloy Ni$_{80}$Fe$_{20}$~at.\% (Py) films and its correlation with microstructure has attracted considerable scientific and industrial interest for decades. The proposed explanations for uniaxial anisotropy include oriented defects and oxides, \cite{sugita1967,fujiwara1968} directional ordering of Fe/Ni atoms pairs, \cite{chikazumi1950a} shape anisotropy of an elongated ordered phase, \cite{kaya1953} composition variation between grains \citep{kench1970} and more recently, localized composition non-uniformity. \citep{rodrigues2018} 
No one of these can account for all instances of uniaxial anisotropy in the Py system, and one or more could contribute simultaneously.

\emph{Epitaxial, single crystal} Py films have a perfect lattice, while the arrangement of Ni and Fe atoms may contain varying degrees of order. Among the explanations above, only the pair ordering and the localized composition non-uniformity are applicable in such a case. Py has vanishingly small magnetocrystalline anisotropy and magnetostriction, and low coercivity, but extremely large magnetic permeability. \citep{chikazumi1961,yin2006} This makes it a unique system in which to study e.g.\ induced uniaxial anisotropy.
Single crystal Py films have been deposited epitaxially by numerous techniques, including thermal evaporation, \citep{chikazumi1961,yelon1965,lo1966} electron beam evaporation, \citep{song1994} molecular beam epitaxy (MBE), \citep{huang1997,tanaka09:2515,tanaka10:345} ion beam  sputtering, \citep{hashim1994,hashim1995} rf magnetron sputtering, \citep{higuchi2011,ohtani2013} dc magnetron sputtering, \citep{michelini2002}  and pulsed laser deposition. \citep{rao2014} 
Most of these deposition methods have resulted in a single crystal Py (001) film on MgO (001) with biaxial anisotropy in the plane, and the easy directions being [110] or [100]. Unfortunately, studies that focused on the growth of single crystal Py did not discuss the effect of ordering at all. Most of these studies used a low deposition rate, which normally results in higher order. On the other hand, the study of atomic order is limited to annealing a quenched specimen at about 500~$^\circ$C for a very long time. \cite{chikazumi1950a,lutts1970,hausmann1971,wan2005}

Several groups have shown that an increase in atomic order results in deterioration of the anisotropy constant $K_1$. \cite{chikazumi1950a,tsukahara1966,hausmann1971,bozorth1993} Uniaxial anisotropy can be induced in single crystal Py by applying an \emph{in-situ} magnetic field during deposition, \citep{chikazumi1961,hashim1994,hashim1995} or by post annealing \citep{bozorth1934b} in magnetic field.
It has been shown that magnetic field induced anisotropy strongly depends on the crystal orientation of the Py for both deposition and annealing in a magnetic field \citep{chikazumi1961,chikazumi1956} i.e.\ it is most efficient along the $\langle111\rangle$ direction, less so along the $\langle110\rangle$ direction and least along the $\langle100\rangle$ direction. 

Another method of inducing uniaxial anisotropy is deposition under an angle with respect to the substrate normal. We have shown that deposition under a 35$^\circ$ angle is more effective than applying a 70~Oe \emph{in-situ} magnetic field when depositing polycrystalline films. \citep{kateb2017,kateb2018}
We have also demonstrated growth of polycrystalline Py films under an angle using high power impulse magnetron sputtering (HiPIMS) and compared with films deposited by conventional dc magnetron sputtering (dcMS). \citep{kateb2018hipims} During HiPIMS deposition high power pulses of low frequency and low duty cycle are applied to a magnetron target which results in highly ionized sputtered material. \cite{gudmundsson12:030801} The HiPIMS discharge provides a highly ionized flux of the metallic species and the averaged ion energy is significantly higher in the HiPIMS discharge than in dcMS discharge and this energetic metallic ions are created during the active phase of the discharge pulse. \cite{bohlmark06:1522,lundin08:035021,greczynski12:4202} For both methods, deposition under an angle with respect to the substrate induces very well-defined uniaxial anisotropy in the film. \citet{schuhl1994} showed that tilt deposition breaks the symmetry between two in-plane easy axes, appearing as a stepped easy axis magnetization loop along the flux direction. However the method of inducing uniaxial anisotropy using tilt deposition of a single crystal Py has not been studied so far. In this work we demonstrate the epitaxial growth of \emph{single crystal} Py films on MgO (001) substrates, by HiPIMS and by dcMS both deposited under an incident angle of 35$^\circ$. We study the effect of the two above mentioned  sputtering methods, whose adatom energy differs by order(s) of magnitude, on the structure, order and magnetic anisotropy of the films. It might be tempting to think that the high adatom energy involved in HiPIMS would cause severe structural damage, but there appear to be only very subtle structural disparities, while the ordering and magnetic anisotropy, however, are vastly different.

\section{Method}

The substrates were single-side polished single crystalline MgO (001) with surface roughness $<$5~\AA, and with lateral dimension of 10$\times$10~mm$^2$ and 0.5~mm thickness (Latech Scientific Supply Pte.~Ltd.). The MgO substrates were used as received without any cleaning but were baked for an hour at 400~$^\circ$C in vacuum for dehydration. The Py thin films were deposited in a custom built ultra-high vacuum magnetron sputter chamber with a base pressure of $<5\times10^{-7}$~Pa. The deposition was performed with argon of 99.999~\% purity as the working gas at 0.33~Pa pressure using Ni$_{80}$Fe$_{20}$ target of 75~mm in diameter and 1.5~mm thickness. During deposition, the substrates were rotated 360$^\circ$ back and forth at $\sim$12.8~rpm with 300~ms stop in between. Further detail on our deposition geometry can be found elsewhere. \citep{kateb2017,kateb2018,kateb2018hipims}


For dcMS deposition a dc power supply (MDX 500, Advanced Energy) was used and the power was maintained at 150~W. For HiPIMS deposition, the power was supplied by a SPIK1000A pulser unit (Melec GmbH) operating in the unipolar negative mode at constant voltage, which in turn was charged by a dc power supply (ADL GS30). 
The pulse length was 250~$\mu$s and the pulse repetition frequency was 100~Hz. The average power during HiPIMS deposition was maintain around 151~W. The HiPIMS deposition parameters were recorded by a home-made LabVIEW program communicating with the setup through high speed data acquisition (National Instruments).

X-ray diffractometry (XRD) was carried out using a X'pert PRO PANalitical diffractometer (Cu K$_\alpha$, wavelength 0.15406~nm) mounted with a hybrid monochromator/mirror on the incident side and a 0.27$^\circ$ collimator on the diffracted side. A line focus was used with a beam width of approximately 1~mm. The film thickness, density and surface roughness was determined by low-angle X-ray reflectivity (XRR) measurements with an angular resolution of 0.005$^\circ$. The data from the XRR measurements were fitted using a commercial X'pert reflectivity program.

For the (002) pole figure the $\theta-2\theta$ was set to a corresponding peak obtained in the normal XRD. However, the (111) and (022) peaks do not appear in the normal XRD. To this end, first a rough pole scan was done according to $\theta-2\theta$ found in the literature. This roughly gives the in-plane ($\phi$) and out-of-plane ($\psi$) angles of those planes with respect to the film surface. Then we scan $\theta-2\theta$ at the right $\phi$ and $\psi$ to find each (111) and (022) peak. Finally, a more precise pole scan is made again at the new $\theta-2\theta$ values. Obviously, the $\theta-2\theta$ values reported in the literature might be slightly different than for our samples due to strain in the film and accuracy of calibration. The $\psi$ was calibrated using the (002) peak of MgO normal to the substrate. In a similar way, the narrow MgO (200) peak in-plane of the substrate was utilized for calibration of $\phi$.

To obtain hysteresis loops, we use a homebuilt high sensitivity magneto-optical Kerr effect (MOKE) looper with HeNe laser light source. We used variable steps in the magnetic field i.e.\ 0.1~Oe steps around transitions of the easy direction, 0.5~Oe steps for the hard axis and before transitions of the easy direction and 1~Oe steps for higher field at saturation.

\begin{figure}[h]
  \centering
  \includegraphics[width=1\linewidth]{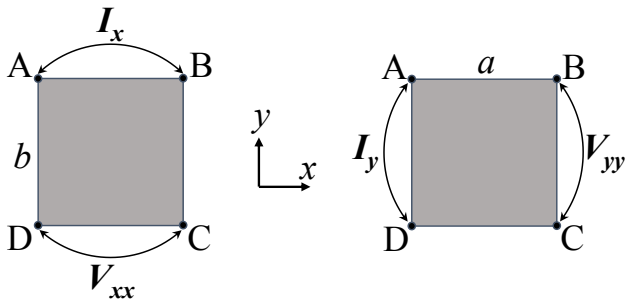}
  \caption{Schematic illustration of set of measurements in vdP method.}
  \label{fig:scheme}
\end{figure}

For the anisotropic magnetoresistance (AMR) measurements we utilized \citet{Price1973} extension to van der Pauw (vdP) \cite{Pauw1958,vdPauw1958} method. We have already shown that the vdP measurement is more reliable in the AMR measurements since it is less geometry dependent compared to conventional Hall-bar method. \cite{kateb2018} Originally, vdP was developed for determining isotropic resistivity and ordinary Hall mobility of uniform and continuous thin films of arbitrary shape and has been used extensively for semiconductor characterization. In the vdP method, four small contacts must be placed on the sample perimeter e.g.\ as illustrated in Fig.~\ref{fig:scheme}. The measured resistances should satisfy vdP equation \cite{vdPauw1958}
\begin{equation}
	\exp\left(-\frac{\pi d}{\rho_{\rm iso}}R_{\rm AB,CD}\right)+\exp\left(-\frac{\pi
	d}{\rho_{\rm iso}}R_{\rm AD,CB}\right)=1
	\label{eq:vdP}
\end{equation}
where $\rho_{\rm iso}$ is the isotropic resistivity and $d$ is the film thickness. The resistance $R_{\rm AB,CD}$ is measured by forcing current through the path ${\rm AB}$ and picking up the voltage at the opposite side between ${\rm CD}$ and $R_{\rm AD,CB}$ is similarly defined. Note that Eq.~(\ref{eq:vdP}) is independent of sample shape and distances between contacts. This behavior is a direct result of conformal mapping i.e.\ a sample has been mapped upon a semi-infinite half-plane with contacts along the edge in which the contact distances cancel out.

It has been shown that the vdP method can be extended to the case of anisotropic films if two of three principle resistivity axis are in the film plane while the 3rd is perpendicular. In that case, $\rho_{\rm iso}$ obtained from Eq.~(\ref{eq:vdP}) stands for the geometric mean of the in-plane principle resistivities i.e.\ $\rho_{\rm iso}=\sqrt{\rho_1\rho_2}$. \cite{hornstra1959,price1972} In principle, the vdP method is based on conformal mapping and thus it should remain valid if an anisotropic medium with lateral dimensions of $a\times b$ mapped into an isotropic one with new dimensions e.g.\ $a\times b'$. In practice one can make a rectangular sample with sample sides parallel to the principle resistivities. \cite{hornstra1959,Price1973} Assuming the principle resistivities are aligned with $x$ and $y$ directions in Fig.~\ref{fig:scheme}, the principle resistivities $\rho_x$ and $\rho_y$ can be obtained from Price's \cite{Price1973} extension, as
\begin{equation}
    \sqrt{\frac{\rho_{x}}{\rho_{y}}}=-\frac{b}{\pi a}\ln\left(\tanh \left[\frac{\pi
    dR_{\rm AD,CB}}{16\rho_{{\rm iso}}}\right]\right)
    \label{eq:rhoratio}
\end{equation}
where $a$ and $b$ are the side lengths of a rectangular sample and $R_{\rm AD,CB}$ is resistance along the $b$ sides as described above. 

Eq.~(\ref{eq:rhoratio}) yields to the ratio of principle resistivities and the individual values can subsequently be obtained by
\begin{equation}
	\rho_{x}=\rho_{\rm iso}\sqrt{\frac{\rho_{x}}{\rho_{y}}}
\end{equation}
and
\begin{equation}
	\rho_{y}=\rho_{\rm iso}\left(\!\sqrt{\frac{\rho_{x}}{\rho_{y}}}\;\right)^{-1} \quad
\end{equation}

For the AMR measurement according to Bozorth's \cite{Bozorth1946} notation one must measure resistivity with saturated magnetization parallel ($\rho_{\|}$) and perpendicular ($\rho_{\bot}$) to current direction. The AMR ratio is given by \cite{McGuire1975}
\begin{equation}
	\rm{AMR}=\frac{\Delta\rho}{\rho_{\rm{ave}}}=\frac{\rho_{\|}-\rho_{\bot}}{\frac{1}{3}\rho_{\|}+\frac{2}{3}\rho_{\bot}}
	\label{eq:amr}
\end{equation}

Assuming the $x$-axis being the current direction, the AMR response can be presented as $\rho_{x}$ which only depends on the direction of saturated magnetization
\begin{equation}
    \rho_{x}=\rho_{\|}+\Delta\rho \cos^2\phi
    \label{eq:rhoxx}
\end{equation}
here $\phi$ stands for angle between current ($x$) and saturated magnetization direction. 

It is worth noting that Eq.~(\ref{eq:rhoxx}) states that the resistivity is only dependent on $\phi$, not on the initial magnetization direction of the films. This is because $\rho_{\|}$ and $\rho_{\bot}$ are measured at saturation magnetization where the entire domains are assumed to be aligned to external magnetic field. \cite{Bozorth1946} 

\section{Results and discussion}
\subsection{HiPIMS discharge waveforms}
\begin{figure}
\includegraphics[width=1\linewidth]{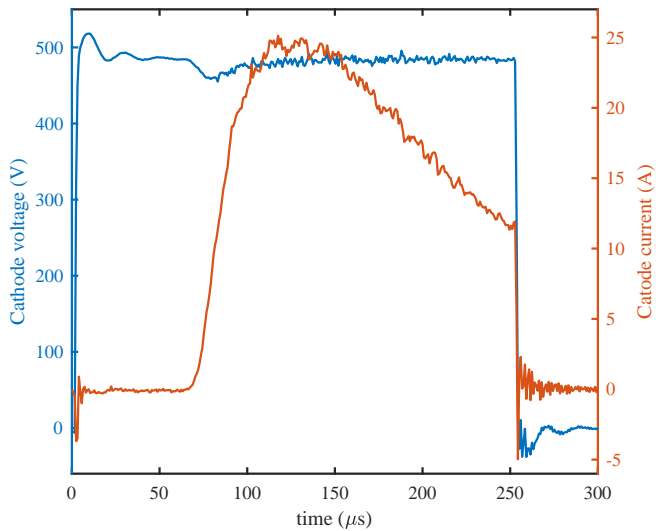}
\caption{\label{fig:waveform} The discharge current and voltage waveforms at 0.33~Pa with the pulse frequency of 100~Hz and pulse length 250~$\mu$s.}
\end{figure}

The discharge current and voltage waveform are a characteristic of the HiPIMS process, which provides important information on both the instrument (the pulser unit) and the physics of ionization. Fig.~\ref{fig:waveform} shows the current and voltage waveforms of the HiPIMS discharge recorded during deposition. It can be seen that a nearly rectangular voltage pulse of 250~$\mu$s length was applied to the cathode target. The oscillations at the beginning and after ending the voltage pulse initiate from internal inductance of the power supply, which creates a resonant circuit along with the capacitance of the cathode target and the parasitic capacitance of the cables. There is also a local minimum corresponding to the current rise at 80~$\mu$s.  

The discharge current is initiated about 70~$\mu$s into the voltage pulse. The dischrge current peaks at 110 $\mu$s into the pulse and then decays until it is cut-off. As described by \citet{lundin09:045008} the current waveforms can be divided into three distinct regions. (I) A strong gas compression due to the rapid accumulation of sputtered flux as plasma ignites which give rise to current to the peak value. This is followed by (II) rarefaction i.e.\ collision of sputter flux with the working gas which results in heating and expansion of the working gas and consequently current decay. More recently, it has been shown that the later mechanism is dominating for rarefaction at higher pressures and ionization loss is dominating otherwise. \cite{huo12:045004} (III) A steady state plasma till the end of the voltage pulse which gives a relatively flat current plateau.

It has been shown that increased pressures can prolong current decay time to the end of pulse and eliminate plateau regions. \cite{lundin09:045008} We believe this is highly unlikely the case in Fig.~\ref{fig:waveform}. We have already shown that 0.33~Pa is in vicinity of minimum pressure i.e.\ lower pressures results in non-linear increase of delay time for current onset and increase time from current onset to peak current which is nearly constant at higher pressures. \cite{kateb2018hipims} Thus pressure is low enough to capture the third stage of the current evolution but, the short pulse length here does not allow it to appear.

It is worth mentinong that, although we have tried to maintain the HiPIMS average power the same as dcMS power (at 150~W), the HiPIMS pulse voltage and peak current (465~V and 25~A) are well above dcMS counterparts (321~V and 463~mA).

\subsection{Microstructure}
\subsubsection{XRR}
Values of film thickness, film density and surface roughness obtained from fitting of the measured XRR curves are shown in Table \ref{XRRfit}. The surface roughnesses obtained here are slightly lower than previously reported values using dcMS. \citep{michelini2002} 
The mass density of HiPIMS deposited film is higher than for the dcMS deposited counterpart. It is somewhat lower than for polycrystalline films deposited by dcMS and HiPIMS (8.7~g/cm$^3$) \cite{kateb2018hipims} and the bulk density of 8.73~g/cm$^3$ (Ref \citenum{mclyman11} (p.~2-6)). Accounting for the epitaxial strain the densities here are within a reasonable range.
\begin{table}[h]
\caption{\label{XRRfit} Values of film thickness, film density and surface roughness obtained by fitting the XRR measurement results.}
\begin{tabular}{ c c c c c }
\hline \hline
Growth & Thickness & Deposition & Density & Roughness\\
technique & (nm) & rate (\AA/s) & (g/cm$^3$) & (\AA) \\
\hline
HiPIMS & 45.74 & 0.97 & 8.38 & 6.75 \\
dcMS & 37.50 & 1.50 & 8.32 & 6.33 \\
\hline
\end{tabular}
\end{table}

It is worth mentioning that, during each HiPIMS pulse, the deposition rate is much higher during the active discharge phase than for dsMS i.e.\ more than 50 times accounting 250~$\mu$s pulse width and 100~Hz frequency.

\subsubsection{XRD}
\begin{figure}
\includegraphics[width=1\linewidth]{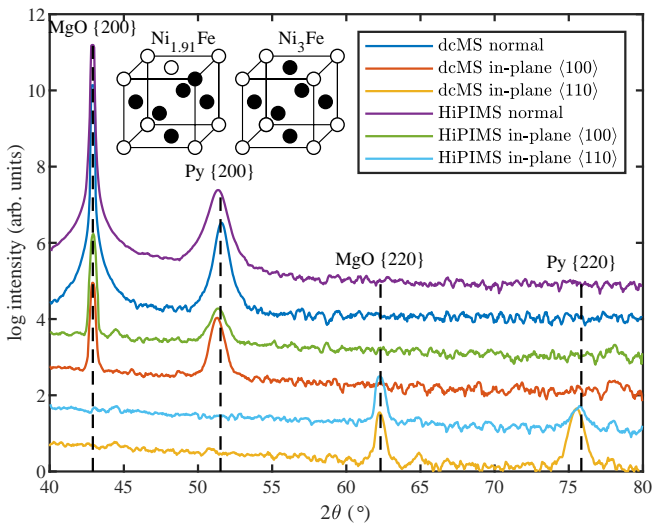}
\caption{\label{XRD} The symmetric XRD pattern of the epitaxial films deposited by HiPIMS (right) and dcMS (left). The vertical dashed lines show the peak position of bulk Py and MgO. The curves are shifted manually for clarity.}
\end{figure}

Permalloy has a fcc structure while both metastable hcp \citep{huang1998,higuchi2011,tanaka10:345} and bcc \cite{yang1999,yin2006,minakawa2016} Py phases have been reported in utrathin films.  Fig.~\ref{XRD} illustrates the symmetric $\theta-2\theta$ XRD pattern of the epitaxial films obtained by both deposition methods. In the out-of-plane XRD, fcc (002) peak is the only detectable Py peak. This indicates that the (002) planes of Py are very well aligned to that of the MgO substrate i.e.\ Py~(001)~$\|$~MgO~(001). Similar results were obtained by measurement in-plane of epitaxial films ($\psi=90^\circ$) along the [100] directions of MgO i.e.\ normal to substrate edges. Furthermore, in-plane measurements along the $\langle$110$\rangle$ direction of MgO (substrate diagonals) show (220) peaks from both the MgO substrate and the Py film. These indicate a orientation relationship of Py~[100]~$\|$~MgO~[100] and Py~[110]~$\|$~MgO~[110] i.e.\ the [100] and [110] directions of Py are fully aligned with those of the MgO substrate. Thus, in spite of the large lattice mismatch (15.84$\%$), high quality single crystal Py ($a_{\rm Py}=3.548$~\AA) film can be established on a MgO ($a_{\rm MgO}=4.212$~\AA) substrate for both deposition techniques. Furthermore, we compared the in-plane peaks along the $\langle100\rangle$ and $\langle010\rangle$ directions (not shown here) and detected no difference in lattice parameter even with a precise scan i.e.\ angular resolution 0.0001$^\circ$ and 10~s counting time. This means we observed identical in-plane strain along the [100] directions in both of the films.

In the dcMS case, the Py (002) peak shows a slight shift towards higher angles in the normal XRD scan. This is accompanied by the shift of in-plane peaks towards smaller angles. Thus, tensile strain at the film-substrate interface generates slight compression normal to the film plane. \cite{tanaka2010} However, in the HiPIMS case, both the in-plane \emph{and} out-of-plane peaks are shifted towards smaller angles. This would indicate tensile strain in all tree dimension that is impossible. However, we attribute the shift of (002) peak in the HiPIMS case to departure from the $L1_2$ Ni$_3$Fe superlattice. \cite{dahl1936} As pointed out by O'Handley (Ref \citenum{Ohandley2000} (p.~548)), the Ni$_3$Fe phase exists in either a disordered or well ordered structure. It has been shown that an ordered Ni$_3$Fe phase can be detected as a shift of XRD peaks towards larger angles \cite{chikazumi1950a,Ohandley2000,wan2005} and narrower peaks. \cite{lutts1970,wan2005} In addition, the intensity of XRD peaks is expected to increase with the higher order. \citep[p.~549]{Ohandley2000} All these conditions can be observed in our dcMS deposited film, indicating that it is more ordered than its HiPIMS counterpart. The more disordered arrangement in the HiPIMS deposition can be attributed to the high deposition rate during each pulse which suppresses adatom mobility. 

\subsubsection{Pole figures}
\begin{figure}
\includegraphics[width=1\linewidth]{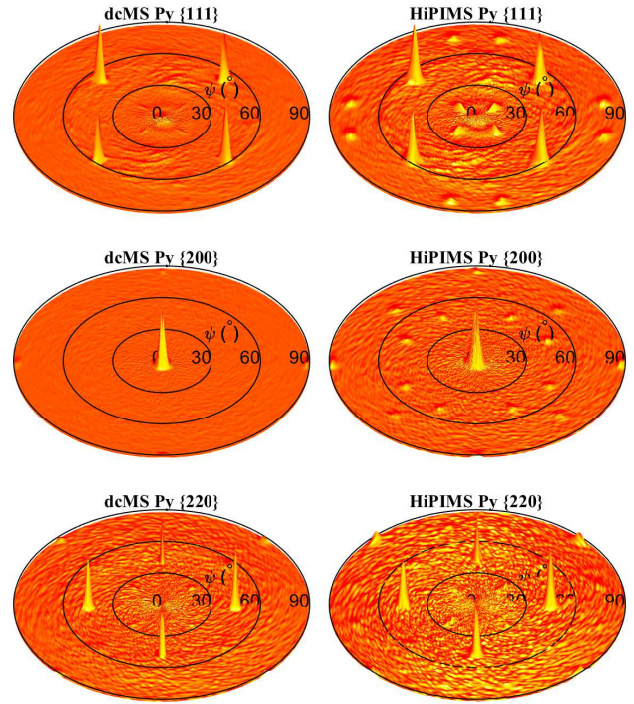}
\caption{\label{Pole} The pole figures obtained for Py \{111\}, \{200\} and \{220\} planes of epitaxial films deposited by HiPIMS and dcMS. The height represents normalized log intensity in arbitrary units.}
\end{figure}

Fig.~\ref{Pole} illustrates pole figures for the main Py planes of our epitaxial films. In the \{200\} pole figure, there is an intense spot at $\psi=0$ that verifies that the (002) plane is lying parallel to the substrate i.e.\ epitaxial relationship of Py(001)~$\|$~MgO(001) for both dcMS and HiPIMS deposited films. There is also a weak spot with four-fold symmetry at $\psi=90^\circ$ due to in-plane diffraction of \{200\} planes parallel to substrate edges in both films. This indicates there are Py \{100\} planes parallel to the substrate edges i.e.\ Py[100]~$\|$~MgO[100]. The \{220\} pole figures, also depict four-fold symmetry of \{220\} planes at $\psi$ angle of 45 and 90$^\circ$ as expected from symmetry in a cubic single crystal for both films. In both of the \{111\} pole figures, there is a four-fold spot at $\phi=45^\circ$ and $\psi=54.74^\circ$ which is in agreement with the angle between (002) and \{111\} planes. However, compared to the (002) spots the \{111\} and \{220\} planes are slightly elongated radially, along the $\psi$ axis. This indicates a lattice constant expanded in-plane of the substrate for both films, in agreement with shift observed in the in-plane XRDs (cf. Fig.~\ref{XRD}). The FWHM of the spots are always narrower for the dcMS deposited epitaxial film indicating higher order in this case.  

The extra dots that appear in the \{111\} pole figure of the HiPIMS deposited film belong to twin boundaries as have also been reported for epitaxially deposited Cu using thermal evaporation \citep{chen2013} and HiPIMS. \citep{cemin2017} The existence of twin boundaries in the Py is a signature of high deposition rate which has been observed previously in evaporated \citep{baltz1963,yelon1965} and electro-deposited \citep{kench1970} films and studied in detail using TEM. \citep{baltz1963,thangaraj1995,ross1996} It can be seen that these dots at 23$^\circ$ also appear in the dcMS deposited film but with very small intensity. This indicates that the fraction of twin boundaries is much lower in the dcMS deposited film. In addition, there are three spots with four-fold symmetry in the \{200\} pole figure of the HiPIMS deposited film which do not appear in the dcMS counterpart. The three dot pattern in the \{200\} pole figure has been characterized as an auxiliary sign of twin boundaries in the film. \citep{cemin2017} It is worth noting that these extra dots in both \{200\} and \{111\} plane were characterized by a $\theta-2\theta$ scan (not shown here) to make sure they belong to the Py film.

\subsection{Magnetic properties}
\begin{figure}
\includegraphics[width=1\linewidth]{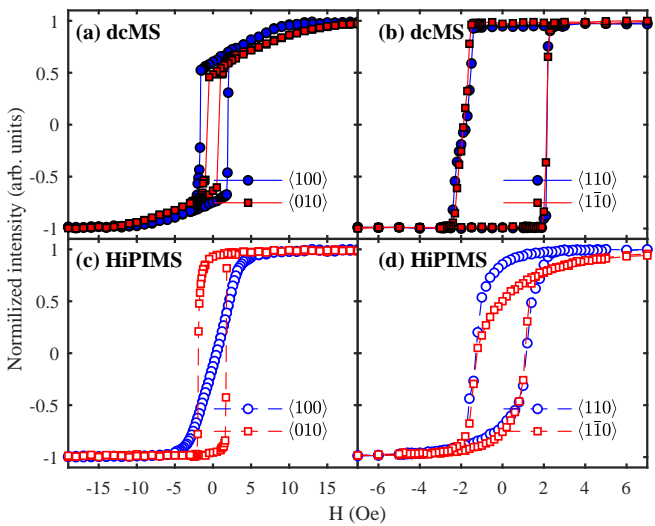}
\caption{\label{MOKE} The average hysteresis loops of the epitaxial films obtained by MOKE measurements along the [100] and [110] directions of the epitaxial Py films.}
\end{figure}

Fig.~\ref{MOKE} compares the results of in-plane MOKE measurements along the [100] and [110] directions of both the epitaxial films. 
Fig.~\ref{MOKE}(a-b) indicate a biaxial behaviour in the dcMS deposited film consisting of two easy axes along the [110] directions with $H_{\rm c}$ of $\sim$2~Oe. This is consistent with the $\langle111\rangle$ direction being the easy direction of the Py crystal and the magnetization being forced in-plane along the $\langle110\rangle$ directions due to shape anisotropy. \citep{yelon1965,ohtake2011} Along the [100] directions the MOKE response is relatively hard  i.e.\ open hysteresis with a gradual saturation outside the hysteresis. The gradual saturation can be explained by an out-of-plane component of the magnetization. \cite{shi2016} In polycrystalline films, the out-of-plane element of magnetization increases with increase in the film thickness \cite{romera2011,silva2017} and it gives perpendicular anisotropy at trans-critical thicknesses. \cite{sugita1967,fujiwara1968,svalov2010} In single crystal films, however, it appears that an out-of-plane component of the magnetization is generally the case. \cite{huang1997,michelini2002,loloee2002}

Fig.~\ref{MOKE}(a) also shows that the $\langle100\rangle$ and $\langle010\rangle$ directions are not completely equivalent for our dcMS deposited film. The $\langle100\rangle$ direction presents larger coercivity ($\sim$2~Oe) and saturates at 12~Oe but the $\langle010\rangle$ direction gives $\sim$1~Oe coercivity and saturates at 15 -- 18~Oe. This difference arises from the fact that $\langle100\rangle$ is the direction of sputter flux during the 300~ms stop time while reversing the rotation. Such a short time is enough to define uniaxial anisotropy in the polycrystalline film using both dcMS and HiPIMS. \cite{kateb2018hipims} However, it appears that for the epitaxial film deposited by dcMS, our deposition geometry is not enough to induce uniaxial anisotropy along the [100] direction in agreement with the previous study of \citet{schuhl1994}. 

As shown in Fig.~\ref{MOKE}(c-d) the HiPIMS deposited epitaxial film shows very well-defined uniaxial anisotropy indicated by a linear hard axis trace without hysteresis and slightly rounded easy axis loop along the [100] directions. 
The anisotropy field ($H_{\rm k}$) of the HiPIMS epitaxial film is 3.5~Oe, i.e.\ much lower than the values observed for polycrystalline films deposited by HiPIMS on Si/SiO$_2$ (11 -- 14.5~Oe). \citep{kateb2018hipims} However the coercivity ($H_{\rm c}$) of 1.8~Oe here is very close to that of polycrystalline films i.e.\ 2 -- 2.7~Oe. We have shown that in polycrystalline films the $H_{\rm c}$ depends on the film density and increases as the film density drops. \cite{kateb2018hipims} In principle the $H_{\rm c}$ of a film depends on the domain boundary structure which has been proved to be dependent on the film thickness. \cite{miyazaki1989} However, since the grain size changes with the film thickness, it is a common mistake to correlate $H_{\rm c}$ with the grain size. We have shown that for a range of film thicknesses (10 -- 250~nm) the grain size changes continuously while $H_{\rm c}$ only changes with the domain wall transition i.e.\ N{\'e}el to Bloch to cross-tie. \cite{kateb2017} 

A question that might arise here is what makes the HiPIMS deposited epitaxial film present uniaxial anisotropy. It has been shown by several groups that formation of ordered Ni$_3$Fe results in lower uniaxial anisotropy constant ($K_1$). \cite{chikazumi1950a,chikazumi1956,tsukahara1966,hausmann1971,bozorth1993} According to both pair ordering \cite{chikazumi1950a} and localized composition non-uniformity \cite{rodrigues2018} theories, uniaxial anisotropy is not expected for a highly symmetric Ni$_3$Fe. While in the case of HiPIMS deposited film, lower order results in uniaxial anisotropy. 

\subsection{Transport properties}

\begin{figure}
\includegraphics[width=1\linewidth]{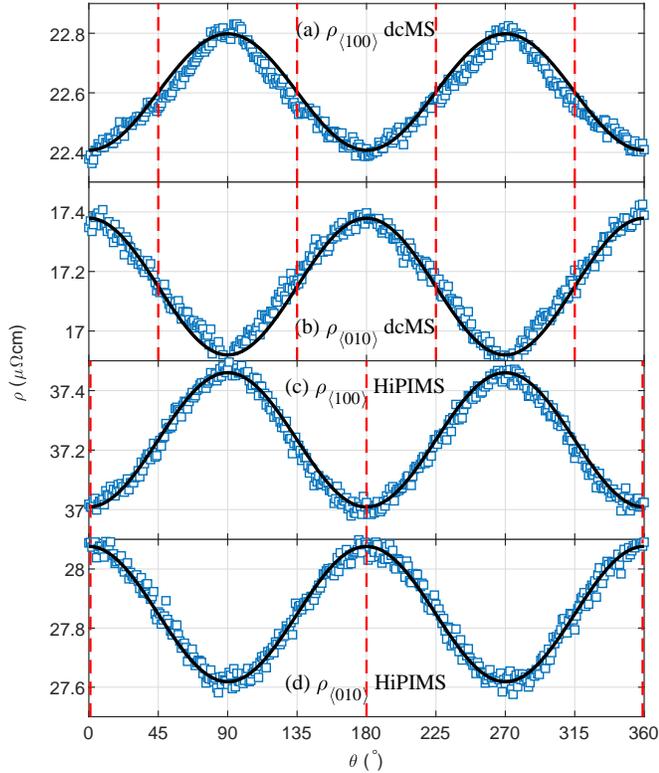}
\caption{\label{fig:amrrot} The AMR obtained by resistivity measurements along the [100] directions of Py films deposited by (a -- b) dcMS and (c -- d) HiPIMS during rotation of 24~Oe magnetic field. The $\theta$ here stands for angle of in-plane magnetization with the $\langle100\rangle$ direction. The black lines indicate the result of fitting with Eq.~(\ref{eq:rhoxx}). The vertical dashed lines indicate the direction of easy axes.}
\end{figure}

Fig.~\ref{fig:amrrot} shows the AMR response of epitaxial films to the rotation of 24~Oe in-plane saturated magnetization. This field is large enough to saturate both films in any direction. The $\theta$ here stands for angle between applied magnetic field and the $\langle100\rangle$ direction of films and should not to be confused with the $\phi$ in Eq.~(\ref{eq:rhoxx}) i.e.\ the angle between current direction and magnetic field. The result of Eq.~(\ref{eq:rhoxx}) is also plotted for comparison as indicated by the black line. Even though, the dcMS deposited film is thinner than the HiPIMS counterpart, the resistivites in the dcMS case are all lower than the HiPIMS ones. This behaviour is in contradiction with the Fuchs model \citep{fuchs1938} which predicts lower resitivity for thicker films. It can be explained in terms of higher Ni$_3$Fe order achieved in the dcMS deposited film. It has been shown previously that the resistivity depends on the order and decreases upon increase in Ni$_3$Fe order. \cite{hausmann1971}

It can be seen that the AMR response of the epitaxial film deposited with HiPIMS conforms better with Eq.~(\ref{eq:rhoxx}) than its dcMS counterpart. In the dcMS case, the deviation from Eq.~(\ref{eq:rhoxx}) occurs at about 45 -- 85$^\circ$, 95 -- 135$^\circ$ and so on. Since the deviation is symmetric around 90$^\circ$ (the $\langle010\rangle$ orientation) it is less likely associated with a pinning mechanism of some domains. Presumably, the deviation originates form switching some domains towards the easy axis at 45 and 135, 225 and 315$^\circ$ i.e.\ [110] orientations. This so-called qusi-static switching in single crystal Py has been studied using torque measurements, as characteristics of biaxial anisotropy. \citep{yelon1965}

The AMR values obtained by Eq.~(\ref{eq:amr}) along the $\langle100\rangle$ and $\langle010\rangle$ directions are summarized in Table~\ref{tab:amr}. We have recently shown that in polycrystalline films the AMR response is different along the hard and easy axis of the film. \cite{kateb2018} It appears that the AMR response is always lower along the $\langle100\rangle$ (direction of flux) in the epitaxial films. It is also evident that higher order reduces resistivity and increases AMR.


\begin{table}[]
    \centering
    \caption{Summary of the AMR results of epitaxial films. (All resitivity values are in $\mu\Omega$-cm unit.)}
    \label{tab:amr}
    \begin{tabular}{c| c c c c c c}
        \hline\hline
        Deposition & Current & $\rho_\|$ & $\rho_\bot$ & $\Delta\rho$ & $\rho_{\rm ave}$ & AMR \\
        method & direction & \multicolumn{4}{c}{($\mu\Omega$-cm)} & (\%) \\
        \hline
        dcMS & $\langle100\rangle$ & 22.19 & 22.80 & 0.39 & 23.06 & 1.70 \\
        dcMS & $\langle010\rangle$ & 16.92 & 16.46 & 0.46 & 16.77 & 2.74 \\
        HiPIMS & $\langle100\rangle$ & 37.46 & 37.91 & 0.45 & 37.76 & 1.19 \\
        HiPIMS & $\langle010\rangle$ & 27.16 & 27.62 & 0.46 & 27.47 & 1.67 \\
        \hline
    \end{tabular}
\end{table}

\section{Summary}
In summary, we have deposited Ni$_{80}$Fe$_{20}$ (001) films by HiPIMS and dcMS. We have characterized them carefully with detailed X-ray measurements, finding only rather subtle structural differences. The pole figures display a signature of twin boundaries (stacking faults) in the HiPIMS deposited film and it appears to be slightly more strained or disordered, regarding dispersion of Ni and Fe atoms, than the dcMS deposited film. However, the differences in the magnetic properties of said films are vast. The dcMS deposited film has biaxial symmetry in the plane, with easy directions [110] as one might expect for a bulk fcc magnetic material (the $\langle111\rangle$ direction is out of plane and shape anisotropy forces magnetization into the plane of the film). The HiPIMS deposited film exhibits different magnetic symmetry, as it has uniaxial anisotropy with $\langle100\rangle$ as the easy direction. Furthermore, the film is magnetically soft and has an anisotropy field of only 3.5~Oe, which is lower than most results we have obtained for polycrystalline films. We attributed the uniaxial anisotropy to less ordered dispersion of Ni and Fe at the atomic level in the film deposited by HiPIMS due to high deposition rate of HiPIMS during the discharge pulse.

\begin{acknowledgments}
The authors would like to acknowledge helpful comments and suggestions from Dr.~Fridrik Magnus and Dr.~Arni S.~Ingason on the structure characterization. This work was partially supported by the Icelandic Research Fund (Rannis) Grant Nos.~196141, 130029 and 120002023, and the Swedish Government Agency for Innovation Systems (VINNOVA) contract No.~2014-04876.
\end{acknowledgments}

\bibliography{Ref}
\end{document}